\documentstyle[axodraw,12pt]{article}          %% 
\makeatletter
\renewcommand{\theequation}{\thesection.\arabic{equation}}
\@addtoreset{equation}{section}
\makeatother

\begin{document}
\vspace*{0.5cm}

\begin{center}

{\large \bf $\bf S, T, U $ parameters 
in  $\bf SU(3)_C\times SU(3)_L\times U(1)$ model\\
 with right-handed neutrinos}
\vspace*{1cm}

{\bf Hoang Ngoc Long}\footnote{Fellow
of the Japan Society for the Promotion of Science }
and {\bf Takeo Inami} 

{\it Department of Physics,
Chuo University,\\
Kasuga, Bunkyo-ku, Tokyo 112-8551, Japan.}

\vspace*{1cm}
\end{center}
The $S,\  T,\ U $  parameters  in the  $ SU(3)_C\times SU(3)_L\times U(1)$ 
model with right -handed neutrinos are calculated.
Explicit expressions for the oblique and 
$Z - Z'$  mixing contributions are obtained. We show that the bilepton
oblique contributions to $S$ and $T$ parameters are bounded :
$- 0.085  \stackrel{<}{\sim} S  \stackrel{<}{\sim} 0.05$~~,
$- 0.001  \stackrel{<}{\sim}  T  \stackrel{<}{\sim} 0.08 $.
The $Z - Z'$ mixing contribution is positive and above 10 \%,
but it will increase fastly with the higher $Z'$ mass. 
%can be negative.
The consequent mass splitting of the bilepton  is  derived and
to be 15 \%. The limit  on the mass of the {\it neutral} 
bilepton in this model is obtained.
 
PACS number(s): 12.60.Cn, 12.10.Dm, 14.80.Cp

\section{Introduction}

\hspace*{0.5cm}Evidence for neutrino oscillation and consequently 
non-zero neutrino mass from the SuperKamiokande atmospheric neutrino 
data are compelling~\cite{suk}. This is the first experimental 
measurement that singnificantly deviates from the standard model (SM), 
and calls for its extension.

Among the possible extensions, the models based on the $\mbox{SU}(3)_C\otimes 
\mbox{SU}(3)_L \otimes \mbox{U}(1)_N$ (3 3 1) gauge
group~\cite{ppf,fhpp} have 
the following intriguing features:
firstly, the models are anomaly free only if the number
of families $N$ is a multiple of three. Further,
from the condition of QCD asymptotic freedom,
which means $N < 5$, it follows that $N$
is equal to 3. The second characteristic
is that  the Peccei-Quinn (PQ)~\cite{pq} symmetry, a solution
of the strong CP problem  naturally occurs in these models~\cite{pal}.
It is worth mentioning that
the implementation of the PQ symmetry
is usually possible only at classical level (it will be  broken
by  quantum corrections through instanton effects),
and there has been a number of attempts to find models
for solving the strong CP question. In the 3 3 1 models the PQ symmetry
following from the gauge invariant Lagrangian  does not
have to be imposed. The third interesting feature
is that one of the quark families  is treated differently
from the other two~\cite{jl,fr95}. This could lead to a natural 
explanation of the unbalancing heavy top quarks in  the fermion
mass hierarchy~\cite{fr95}. Recent analyses have indicated that signals
of new particles in this model, bileptons~\cite{can} and exotic
quarks~\cite{jm} may be observed at the Tevatron and the Large
Hadron Collider (LHC).\par
\bigskip

There are two main versions of the 3 3 1 models: the minimal in which
all lepton components ($\nu, l, (l_R)^c$) belong to the
same lepton triplet and a variant, in which right-handed neutrinos 
(r. h. neutrinos) are included i.e. $\nu, l, \nu_L^c$ are in the triplet
(hereafter we call it a model with right-handed neutrino~\cite{rhnm,mpp}).
New gauge bosons in the minimal model are  bileptons
 ($Y^\pm, Y^{\pm\pm}$) carrying lepton number $L = \pm 2$ and $Z'$.
Most analyses of the 3 3 1 models have centred on the 
bileptons~\cite{saken,frha,cuda,rev} and $Z'$~\cite{zi,jj}.
In the second model, the bileptons with lepton number $L = \pm 2$
are singly-charged $Y^\pm$ and {\it neutral} gauge bosons  
$X^0, \bar{X}^0$ , and both are responsible for lepton violating
interactions. This model is interesting
because of the existence of r.h. neutrinos and the neutral bilepton
$X^0$, the later being a promissing candidate in accelerator 
experiments~\cite{rgh}. 
Since the symmetry of the $SU(2)_L$ gauge group
is broken, generically the  neutral bilepton 
has a mass $M_{X^0}$  different from that of the
singly-charged bilepton $M_{Y^+}$.  However, looking at recent
review~\cite{cuda}  we see that  there is almost no bound on the 
$X^0$ mass (the limit given there  for $M_{X^0}$ is above 44 GeV).

%On the other hand it is known that the 
Heavy particles can be indirectly observable via radiative corrections
in the SM-type theories~\cite{itl}.
At present the oblique radiative parameters $S, T$~\cite{pt},
and $U$~\cite{hold}
can be used optimally to extract new-physics effects.
In the early papers
the focus was on fermionic contributions~\cite{takeo}. 
The aim of this paper is to calculate the $S,\  T,\  U$ parameters,
and to get a bound on the bilepton masses.

This paper is organized as follows:
In Sec. 2 we briefly introduce necessary  elements of the 
model, and the bilepton mass splitting due to the symmetry
breaking is given.  
 Sec. 3 is devoted to calculating the new gauge boson contributions
to the $S,\ T,\ U$ parameters. We make a remark on
the minimal model in Sec. 4. A numerical evaluation is presented in 
Sec. 5. We summarize our result and make conclusions  
in the last section.

\section{The model and bilepton mass splitting }

\hspace*{0.5cm}In this section we firstly  recapitulate the basic 
elements of the model. Based on the VEV structure and the muon decay 
experiment we obtain a bound on the neutral bilepton mass $M_{X^0}$.
The details  can be found  in~\cite {rhnm}.
In the variant of the 3 3 1 model the third member  of the  lepton 
triplet  is r.h. neutrino instead of the antilepton $l^c_L$
\begin{equation}
f^{a}_L = \left( 
               \nu^a_L,\  e^a_L,\ (\nu^c_L)^a
                 \right)^T \sim (1, 3, -1/3),
\label{l}
\end{equation} 
where a = 1, 2, 3 is the generation index. 

This assignment  leads to   the electric charge and hypercharge operators
which are now defined by
\[ Q=\frac{1}{2}\lambda_3-\frac{1}{2\sqrt{3}}\lambda_8+N,\
 Y=2N-\lambda_8/\sqrt{3}, \hspace*{0.3cm} (\lambda_8= {\rm diag}(1, 1, 
-2)/\sqrt{3}).\]
 
The exotic quarks have charges 2/3 and -- 1/3 
 and are $SU(2)_L$ singlets
\begin{eqnarray}
 Q_{iL} & = & \left( \begin{array}{c}
                d_{iL}\\-u_{iL}\\ D_{iL}\\ 
                \end{array}  \right) \sim (3, \bar{3}, 0),
\   D_{iR}\sim (3, 1, -1/3),\ i=1,2, \nonumber\\
 Q_{3L}& = & \left( \begin{array}{c}
                 u_{3L}\\ d_{3L}\\ T_{L}
                \end{array}  \right) \sim (3, 3, 1/3),
\   T_{R}\sim (3, 1, 2/3). 
\end{eqnarray}

The symmetry breaking can be 
achieved with  three $SU(3)_{L}$ Higgs triplets
\begin{equation}
\chi = \left(
                \chi^o,\ \chi^-,\ \chi^{,o} 
                  \right)^T,\
\rho = \left( 
                \rho^+,\ \rho^o,\ \rho^{,+} 
                 \right)^T ,\
\eta = \left( 
                \eta^o,\ \eta^-,\ \eta^{,o} 
                  \right)^T.
\label{h2}
\end{equation}
They acquire the  vacuum expectation values (VEVs):
$\langle\chi \rangle^T = (0, 0, \omega/\sqrt{2})$,\
$\langle\rho \rangle^T = (0, u/\sqrt{2}, 0)$,\  and
$\langle\eta \rangle^T = (v/\sqrt{2}, 0, 0)$.
The gauge symmetry is broken to the SM gauge 
symmetry by $\omega \neq 0$.

The complex gauge bosons $\sqrt{2}\ W^+_\mu = W^1_\mu-iW^2_\mu$ ,$
\sqrt{2}\ Y^-_\mu = W^6_\mu- iW^7_\mu ,\sqrt{2}$\ $ X^0_\mu =
W^4_\mu- iW^5_\mu $  have the following masses
\begin{equation}
m^2_W=\frac{1}{4}g^2(u^2+v^2), M^2_Y=\frac{1}{4}g^2(v^2+\omega^2), 
M^2_X=\frac{1}{4}g^2(u^2+\omega^2).
\label{mnhb}
\end{equation}
The physical neutral gauge bosons are mixtures of  $Z,\ Z'$:
\begin{eqnarray}
Z^1  &=&Z\cos\phi - Z'\sin\phi,\nonumber\\
Z^2  &=&Z\sin\phi + Z'\cos\phi.
\end{eqnarray}
Here the photon field $A_\mu$ and $Z,Z'$ are given by~\cite{rhnm}:
\begin{eqnarray}
A_\mu  &=& s_W  W_{\mu}^3 + c_W\left(-\frac{t_W}{\sqrt{3}}\ W^8_{\mu}
+\sqrt{1-\frac{t^2_W}{3}}\  B_{\mu}\right),\nonumber\\
Z_\mu  &=& c_W  W^3_{\mu} - s_W\left(-\frac{t_W}{\sqrt{3}}\ W^8_{\mu}+
\sqrt{1-\frac{t_W^2}{3}}\  B_{\mu}\right), \nonumber \\
Z'_\mu &=& \sqrt{1-\frac{t_W^2}{3}}\  W^8_{\mu}+\frac{t_W}{\sqrt{3}}
\ B_{\mu},
\label{apstat}
\end{eqnarray}
where the usual notation is used: $s_W \equiv \sin\theta_W$.
The mixing angle $\phi$ is given by
\begin{equation}
\tan^2\phi =\frac{m_{Z}^2-m^2_{Z^1}}{M_{Z^2}^2-m_{Z}^2},
\label{tphi}
\end{equation}
where $m_{Z^1}$ and $M_{Z^2}$ are the {\it physical} mass eigenvalues with
\begin{eqnarray}
m_{Z}^2   &=&\frac{g^2}{4 c_W^2}(u^2+v^2)=\frac{m_W^2}{c_W^2}, 
\label{masszw}\\
M_{ZZ'}^2&=&\frac{g^2}{4c_W^2\sqrt{3-4s_W^2}}
\left[u^2-v^2(1-2s_W^2)\right],
\label{mzzp} \\
M_{Z'}^2 &=&\frac{g^2}{4(3-4s_W^2)}\left[4\omega^2+ \frac{u^2}{c_W^2}
+ \frac{v^2(1-2s_W^2)^2}{c_W^2}\right].
\label{masmat}
\end{eqnarray}

One of the  Higgs bosons can be identified with the SM
Higgs~\cite{hnl98}.

The  lower limit on the singly-charged
bilepton is obtained by the ``wrong''
muon  decay ~\cite{rpd}
\begin{equation} R = \frac{\Gamma (\mu^- \rightarrow e^- \nu_e \bar{\nu}_\mu)}
{\Gamma (\mu^- \rightarrow e^- \bar{\nu}_e \nu_\mu)}
\sim \left(\frac{m_W}{M_Y}\right)^4.
\label{lowlm}
\end{equation}
The observed limit $R < 1.2\% $  (at  90\%  CL) gives
 $M_{Y^-} \ge 230$ \ GeV.

From~(\ref{mnhb}) we get a bound on the bilepton mass splitting
\begin{equation}
| M^2_Y -  M^2_X | \leq m_W^2.
\label{masplit}
\end{equation}
Combining~(\ref{lowlm}) and~(\ref{masplit}) we get the {\it first}
prelimilary constraint on the neutral bilepton mass:
\begin{equation}
M_{X^0} \geq 230 \pm 17\  {\rm GeV},    \ 90\%\   {\rm CL}.
\label{first}
\end{equation}

In conclusion, the model predicts three kinds of new particles:
new gauge bosons $Y^\pm $, $ X^0 $,$\bar{ X}^{o}$ and $Z'$, new exotic quarks 
$T$, $ D_1$, $ D_2$ and new Higgs scalars. Bileptons ($Y^+, X^0$) make
an $SU(2)_L$ doublet with hypercharge $Y = 1/2$, while exotic quarks
and $Z'$ are $SU(2)_L$ singlets. Due to the VEV structure, the mass
splitting of the bileptons is bounded by the SM $W$ boson mass $m^2_W$.

\section{ Contributions of new particles to
  $\bf S,\ T,\ U$ parameters}

\hspace*{0.5cm}Since new quarks  are $SU(2)_L$ singlets, they do 
not enter into  the oblique corrections to the $S,\ T,\  U$ parameters
which are only sensitive to $SU(2)$ 
breaking. Similarly,
$Z'$ will not contribute except through $Z - Z'$ mixing.

 {\bf 3.1 Effective interaction}\\[0.3cm]
\hspace*{0.5cm}We begin by writing the Lagrangian for the bileptonic 
gauge field $Y$
and the Higgs field $\Phi$ below the $SU(3)_L$ breaking scale.
They are  $SU(2)_L$ doublets with the hypercharge $Y = \frac{1}{2}$
\begin{equation}
\Phi = \left(
\begin{array}{c}
\Phi^+ \\ \Phi_0
\end{array}
\right),\
Y_\mu = \left(
\begin{array}{c}
Y^+_\mu \\ X^0_\mu
\end{array}
\right).
\end{equation}

The effective Lagrangian is given

\begin{eqnarray}
{\cal L}_{0} &=& - \frac{1}{2} (Y_{\mu\nu})^{\dag} Y^{\mu\nu}
+ (D_\mu \Phi - i M Y_\mu)^{\dag} (D^\mu \Phi - i M Y^{\mu})
\nonumber\\
&& 
- i  g Y_\mu^{\dag} F^{\mu\nu}(W) Y_\nu
+ i \frac{1}{2}  g' Y_\mu^{\dag} F^{\mu\nu}(B) Y_\nu
\ ,
\label{LY0}
\end{eqnarray}
where $M$ is $2\times2$ matrix given by
\begin{equation}
M = \left(
\begin{array}{cc}
M_+ & 0 \\ 0 & M_0
\end{array} \right)
\ ,
\end{equation}
and  $D_\mu  =  \partial_\mu - i g W_\mu + i \frac{1}{2} g' B_\mu $
with  $g = \sqrt{3} g'$.
For the shorthand hereafter we denote $M_{Y^+} \equiv M_+,\  M_{X^0} 
\equiv M_0$.
%%%%%%%%%%%%%%%%%%%%%%%%%
\begin{figure}
\begin{center}
\begin{picture}(360,320)(0,0)
\Text(60,290)[]{(a)}
\PhotonArc(110,275)(25,0,360){2}{20}
\Photon(70,275)(83,275){2}{2}
\Text(70,265)[]{I}
\Photon(135,275)(150,275){2}{2}
\Text(145,265)[]{J}
\Text(240,290)[]{(b)}
\PhotonArc(290,275)(25,0,360){2}{20}
\Photon(290,250)(265,240){2}{3}
\Photon(290,250)(315,240){2}{3}
\Text(270,233)[]{I}
\Text(310,233)[]{J}
\Text(60,205)[]{(c)}
\DashArrowArc(116,190)(25,0,180){2}{3}
\DashArrowArc(110,190)(25,180,360){2}{3}
\Photon(68,190)(80,190){2}{2}
\Photon(131,190)(146,190){2}{2}
\Text(68,180)[]{I}
\Text(146,180)[]{J}
\Text(240,205)[]{(d)}
\DashCArc(290,190)(25,0,360){2}{10}
\Photon(240,190)(255,190){2}{2}
\Photon(305,190)(320,190){2}{2}
\Text(240,180)[]{I}
\Text(320,180)[]{J}
\Text(40,130)[]{(e)}
\DashCArc(90,115)(25,0,360){2}{20}
\Photon(84,90)(62,80){2}{3}
\Photon(84,90)(105,80){2}{3}
\Text(62,70)[]{I}
\Text(105,70)[]{J}
\Text(220,130)[]{(f)}
\PhotonArc(270,115)(25,180,360){2}{12}
\DashCArc(270,115)(25,0,180){2}{20}
\Photon(220,115)(235,115){2}{2}
\Photon(285,115)(300,115){2}{2}
\Text(220,105)[]{I}
\Text(300,105)[]{J}
\label{fig:pinch}
\end{picture}
\end{center}
\caption[]{ Feynman diagrams contributing 
to  vacuum polarizations  $\Pi_{IJ}$ (I, J = 1, 3, 8). 
Wavy lines denote bileptons $X,Y$, dashed lines  associated 
 WbNG bosons, and arrow dashed   FP ghosts.}\label{loop} 
\end{figure}
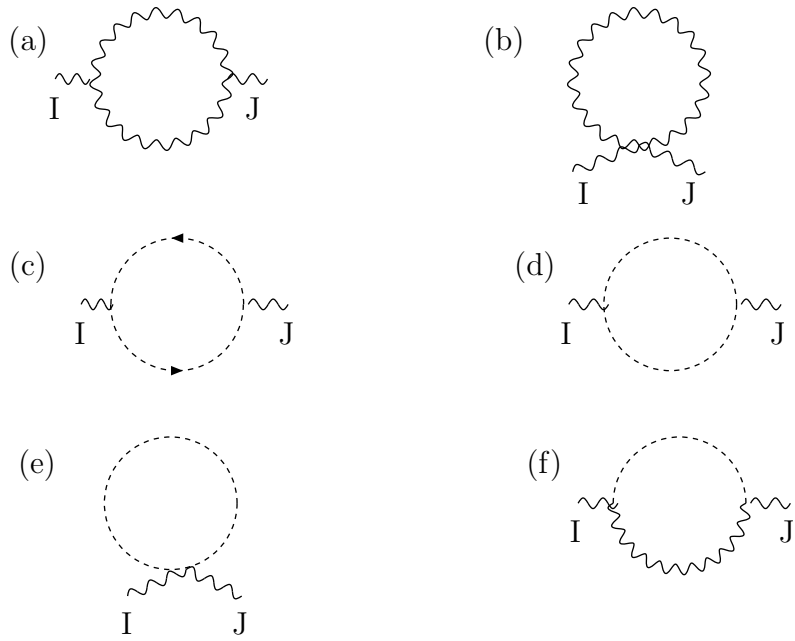
%%%%%%%%%%%%%%%%%%%%%%%%%

{\bf 3.2 Oblique corrections}\\[0.3cm]
\hspace*{0.5cm}As it was shown in Ref.~\cite{saken} contributions from 
Higgs fields turn on   the masses of gauge bosons:
the SM, $Z'$ and  bileptons.
One loop diagrams contributing to vacuum polarizations $\Pi_{IJ}(I,J =
1, 3, 8)$  are
shown in Figure 1. The diagram (c) is the Faddeev-Popov (FP) 
ghost contribution and the diagrams  (d) -- (f) are the contributions 
of the  WbNG bosons.
The calculations below were done 
in the  't Hooft-Feynman gauge. 
For convenience we use the following functions which will arise 
when a bilepton of one  kind
and its associated FP ghost and WbNG boson go around the loops.
\begin{eqnarray}
E(q^2,M^2)&=&q^2[3\Delta+\frac{2}{3}- \bar{F}_0(q^2,M,M)-12 
  \bar{F}_3(q^2,M,M)-3\ln M^2]\nonumber\\
& & \qquad  + M^2(\Delta  + 1 - \ln M^2),\nonumber\\
E'(q^2,M^2)&=&q^2[3\Delta+\frac{2}{3}- \bar{F}_0(q^2,M,M)-12 
  \bar{F}_3(q^2,M,M)-3\ln M^2]\nonumber\\
& & \qquad  - 2 M^2(3\Delta + 1 - 3\ln M^2 ).\nonumber
\end{eqnarray}
Here 
\[  \Delta \equiv \frac{2}{4-n} - \gamma_E - \ln(\pi) \]
where $n$ is the space-time dimensionality, and $ \gamma_E $
is the Euler-Mascheroni constant.

The vacuum polarizations   are  then summarized by
\begin{eqnarray}
\Pi_{38} & = & \frac{1}{64 \pi^2 \sqrt{3}}
[E(q^2,M_0^2) - E(q^2,M_+^2)]\nonumber\\
\Pi_{33} & = & \frac{1}{64 \pi^2}
[E'(q^2,M_+^2) + E'(q^2,M_0^2)]\\
\Pi_{11} & = & \frac{1}{32 \pi^2} 
\Biggl\{
 q^2[3\Delta - \frac{2}{3} - 5  \bar{F}_0(q^2,M_+,M_0) + 12 
   \bar{F}_3(q^2,M_+,M_0)-3\ln(M_+M_0)]\nonumber\\
& &  - M^2_+[3 \Delta +1 -
   \bar{F}_0(q^2,M_+,M_0) ] -  M^2_0[3 \Delta +1 -
   \bar{F}_0(q^2,M_+,M_0) ] \nonumber\\
& & - 10   \bar{F}_4(q^2,M_+,M_0) + 5(M^2_+\ln M^2_++M^2_0\ln M^2_0)
\nonumber\\ 
& &\qquad - 2(M^2_+\ln M^2_0+M^2_0\ln M^2_+) 
\Biggr\},\nonumber
\end{eqnarray}
where functions $  \bar{F}_0(s,M,m),\   \bar{F}_3(s,M,m)$ and 
$  \bar{F}_4(s,M,m)$  are defined in
Ref.~\cite{frha}. They differ from those $F$s in~\cite{saken}
by a term proportional  to  $\ln (Mm) $. 

For later use we write down the mentioned
functions at $q^2 = 0$  and  small $q^2$ behavior:
\begin{eqnarray}
  \bar{F}_0(0,M,m)& =& - \frac{1}{2} \left[ 2 -  
\frac{ M^2 + m^2}{ M^2 - m^2 } \ln \frac{M^2}{m^2}  \right]\nonumber\\
  & =  & - \frac{\varepsilon^2(M,m)}{4} +  O( \varepsilon^3(M,m)),
\nonumber\\
  \bar{F}_3(0,M,m)& =& -\frac{5}{36} + \frac{M^2 m^2}{3(M^2 - m^2)^2} +
\frac{M^2 + m^2}{12 ( M^2 - m^2)}\left[ 1 - 
\frac{2 M^2 m^2}{(M^2 - m^2)^2}\right]
  \ln \frac{M^2}{m^2} \nonumber\\
 & =  &  \frac{\varepsilon^2(M,m)}{12} +  O( \varepsilon^3(M,m)),
\nonumber\\
  \bar{F}_4(0,M,m) & = & - \frac{1}{4}\left[ M^2 + m^2 -
\frac{M^4 + m^4}{M^2 - m^2}\ln \frac{M^2}{m^2}\right]\nonumber\\
 & =  &  m^2\left[ \frac{\varepsilon^2(M,m)}{6} +  O( \varepsilon^3(M,m))
\right],
\label{Ffunction}\\
 \bar{F}_A(0,M,m) & = & \frac{1}{2(M^2 - m^2)}\left[ M^2 + m^2 -
\frac{2 M^2 m^2}{M^2 - m^2}\ln \frac{M^2}{m^2}\right]\nonumber\\
 & =  & \frac{\varepsilon(M,m)}{6}\left[ 1 -  
\frac{\varepsilon(M,m)}{2} \right] +  O( \varepsilon^3(M,m)),
\nonumber
\end{eqnarray}
where $\varepsilon(M,m) \equiv \frac{M^2 - m^2}{m^2}$.
In the case of identical masses $ m = M $  we have
\begin{eqnarray}
 \bar{F}_0(m^2_Z,M,M)& = & - \frac{\delta (M)}{6}\left[ 1 +
 \frac{\delta (M)}{10}  +  \frac{\delta^2 (M)}{70} \right] + 
O( \delta^4 (M)),\nonumber\\
    \bar{F}_3(m^2_Z,M,M)& = & \frac{1}{6}\left[ 1 + 
\frac{2 M^2}{m^2_Z}\right] \bar{F}_0(m^2_Z,M,M) + \frac{1}{18},
\label{fd}
\end{eqnarray}
where $\delta(M) \equiv \frac{m^2_Z}{M^2}$.
The function  $\bar{F}_4$ can be calculated through  
$\bar{F}_0$  by the following relation:
\begin{eqnarray}\bar{F}_4(m^2_W,M,m)& = &  \frac{M^2 + m^2}{2} 
\bar{F}_0(m^2_W,M,m) \nonumber\\
 & & -
 \frac{M^2 - m^2}{2 m^2_W}\left[ \bar{F}_0(m^2_W,M,m) - 
 \bar{F}_0(0,M,m) \right].
\label{f4}
\end{eqnarray}
Other  useful formulas are given in Appendix A of Ref.~\cite{frha}.

The contributions to $S,\ T, \ U$  from bileptonic gauge bosons
coming through the  transverse self-energies are given
\begin{eqnarray}
S_{\rm pol}& = & - 16 \pi {\rm Re} \frac{ \Pi^{3Y}(m_Z^2) - 
\Pi^{3Y}(0) }{m_Z^2} \nonumber\\
&=& \frac{1}{4\pi} 
\Biggl\{\ln \frac{M_{+}^2}{M_0^2} + \frac{1}{3} 
  \left[   \bar{F}_0(m_Z^2,M_+,M_+) -   
\bar{F}_0(m_Z^2,M_0,M_0)\right]\nonumber\\
& &\qquad + 4 \left[  \bar{F}_3(m_Z^2,M_+,M_+) -   
\bar{F}_3(m_Z^2,M_0,M_0)\right]
\Biggr\},\nonumber\\
T_{\rm pol}& =& \frac{4\sqrt{2}G_F}{\alpha}
\left( \Pi^{11}(0) - \Pi^{33}(0) \right)\nonumber\\
& = &\frac{3\sqrt{2}G_F}{16\pi^2\alpha}
\left[ M_+^2 + M_0^2 - \frac{2M_+^2M_0^2}{M_+^2-M_0^2} 
\ln \frac{M_+^2}{M_0^2}\right],
\label{tnp}
\\
U_{\rm pol} &=& 16\pi \left[
  \frac{\Pi^{11}(m_W^2) - \Pi^{11}(0)}{m_W^2}
  - \frac{\Pi^{33}(m_Z^2) - \Pi^{33}(0)}{m_Z^2}\right]\nonumber\\
&=& - \frac{1}{\pi}
\Biggl\{\frac{2}{3} - \frac{(M^2_+ + M^2_0)}{2\ m_W^2}
  \left[  \bar{F}_0(m_W^2,M_+,M_0) -   
\bar{F}_0(0,M_+,M_0)\right] \nonumber\\
& & - \frac{1}{4}\left[  \bar{F}_0(m_Z^2,M_+,M_+) +   
\bar{F}_0(m_Z^2,M_0,M_0) \right]
 + \frac{5}{2}   \bar{F}_0(m_W^2,M_+,M_0)\nonumber\\
& &    - 3 \left[  \bar{F}_3(m_Z^2,M_+,M_+) + 
  \bar{F}_3(m_Z^2,M_0,M_0) \right]\nonumber\\
& & + \frac{5}{m_W^2}  \left[  \bar{F}_4(m_W^2,M_+,M_0) - 
      \bar{F}_4(0,M_+,M_0)\right] - 6 \bar{F}_3(m_W^2,M_+,M_0)
\Biggr\}.\nonumber
\end{eqnarray}
  
It is known that the bosonic contributions to the $S, T $, and $U$ 
parameters defined in
terms of conventional self-energies, are gauge dependent
and, moreover, divergent unless the restrictive condition~\cite{pin}
\[ \xi_W = c^2_W \xi_Z + s^2_W \xi_\gamma.\]
is imposed.
The parameters become  gauge invariant after adding the  
pinch parts arising from vertex and box diagrams.

The self-energies of electroweak gauge bosons are modified by
pinch parts which can be expressed as~\cite{pin}
\begin{eqnarray}
\left. \Pi_{ZZ}(q^2) \right\vert_P & = &- (q^2-m_Z^2) 
\Bigl[ B_0(q^2,M_0,M_0) \nonumber\\
&&\qquad + (1-2 s^2_W)^2 B_0(q^2,M_{+},M_{+})\Bigr],\nonumber\\
\left. \Pi_{ZQ}(q^2) \right\vert_P &=&
- (2q^2-m_Z^2) (1-2 s^2_W) B_0(q^2,M_+,M_+),
\label{pinsf}\\
\left. \Pi_{QQ}(q^2) \right\vert_P &=&
- 4q^2 B_0(q^2,M_+,M_+)\ ,\nonumber\\
\left. \Pi_{WW}(q^2) \right\vert_P &=&
- 2(q^2-m_W^2) B_0(q^2,M_+,M_0) \ ,\nonumber
\end{eqnarray}
where $B_0$ is defined by
\begin{eqnarray}
B_0(q^2,M_1,M_2)& =& \int \frac{d^nk}{i(2\pi)^n}
\frac{1}{[M_1^2-k^2][M_2^2-(k+q)^2]} \nonumber\\
& = & \frac{1}{16 \pi^2} \left[ \Delta +
\ln(M_1 M_2) +  \bar{F}_0(q^2,M_1,M_2)\right]\  .\nonumber
\end{eqnarray}
In getting (\ref{pinsf}) we have used coupling constants
of bileptons $X,\ Y$ with the SM vector bosons: the photon $A$,
weak-bosons  $Z$ and  $W$. In the  notations of Ref.~\cite{grace} 
they are given by
\begin{eqnarray}
CAYY &=&g\ s_W, \ CAXX=0, \ CZXX = - \frac{g}{2 \ c_W},
\nonumber\\
CZYY & = & \frac{g(1-2\ s^2_W)}{2 \ c_W},\
CWXY = \frac{g}{\sqrt{2}}.\nonumber
\end{eqnarray}

The above pinch parts  give the following  corrections 
to  $S,\  T$  and  $U$ parameters~\cite{kl} 
\begin{eqnarray}
S_{\rm pin} & = &\frac{16\pi}{m_Z^2}{\rm Re}
\Biggl[ \Pi_{ZZ}(m^2_Z)-\Pi_{ZZ}(0) -
 (1-2s_W^2)( \Pi_{ZQ}(m^2_Z)-\Pi_{ZQ}(0) )\nonumber\\
& & \qquad - s^2_W(1-s^2_W)\Pi_{QQ}(m^2_W)\Biggr]\nonumber\\
&=& \frac{1}{\pi}
\Biggl[ \ln \frac{M_+^2}{M_0^2}  +   \bar{F}_0(m_Z^2,M_+,M_+)\Biggr],
\nonumber\\
T_{\rm pin} & = &\frac{4\pi}{s^2_W c^2_W}{\rm Re}
\Biggl[ c^2_W \frac{\Pi_{WW}(0)}{m^2_W} - \frac{1}{m^2_Z}
\left(\Pi_{ZZ}(0) + 2 s^2_W \Pi_{ZQ}(0) \right)\Biggr]\nonumber\\
& =& \frac{1}{4\pi\ s^2_W}
\left[ 2   \bar{F}_0(0,M_+,M_0) +  \ t^2_W \ln\frac{M_+^2}{M_0^2}
\right],
\label{tpin}\\
 U_{\rm pin}  & = &16 \pi{\rm Re}
\Biggl\{ \frac{ \Pi_{WW}(m^2_W)-  \Pi_{WW}(0)}{m^2_W} + \frac{1}{m^2_Z}
\left[ \Pi_{ZZ}(0)\right. \nonumber\\
& & \qquad \left. - 2 s^2_W( \Pi_{ZQ}(m^2_Z)-\Pi_{ZQ}(0) )
   - s^4_W \Pi_{QQ}(m^2_Z)\right]\Biggr\}\nonumber\\
&=& \frac{2}{\pi}
\Biggl[ \ s_W^2   \bar{F}_0(m_Z^2,M_+,M_+) -   
\bar{F}_0(0,M_+,M_0)\Biggr]\ .
\nonumber
\end{eqnarray}

The expression $S_{\rm pin},\ T_{\rm pin}$ and $U_{\rm pin}$  in
Eq.(\ref{tpin})  must be added to    $S_{\rm pol},\  
T_{\rm pol}$  and   $U_{\rm pol}$  
in Eq.(\ref{tnp}). Note that due to the term proportional to
$\ln\frac{M_+^2}{M_0^2} $  the pinch parts can give negative
contributions to the $S$ and $T$ parameters. It is well known that 
the oblique parameter $T$ is positive both for the case of a heavy
left-handed fermion doublet and for the case of a scalar doublet
of general hypercharge. On the other hand, the present experimental
data seem to favor a negative value for $T$. So our model is good in 
that sense.
 
{\bf 3.3 The $Z - Z'$ mixing contribution}\\[0.3cm]
\hspace*{0.5cm}The effects of the $Z-Z'$ mixing in
 a general context  has been considered in~\cite{zprim}.

Now, due to the $Z - Z'$ mixing, the observed 
$Z$ boson mass $m_{Z_1}$ at LEP1 or SLC is shifted from 
the SM $Z$ boson mass $m_Z$: 
\begin{equation}
\Delta m^2 \equiv m_{Z_1}^2 - m_Z^2  = - \tan^2 \phi\left( M^2_{Z_2}
- \frac{m^2_W}{c^2_W}\right) \le 0. 
\label{masshif}
\end{equation}
In writing down the last equality of Eq.~(\ref{masshif}),
we have employed Eq.~(\ref{masszw}).

The presence of the mass shift affects the 
$T$-parameter  at tree level~\cite{pt,hold}. 
The result is~\cite{hold}:
\begin{equation}
T_{zz'}  =  -\frac{ \Delta m^2}{\alpha\  m^2_{Z_1}}\
         =  \frac{\tan^2 \phi}{\alpha}\left( \frac{
M^2_{Z_2}}{m^2_{Z_1}} - \frac{m^2_W}{c^2_W m^2_{Z_1}} \right)        
     \simeq  \frac{\tan^2 \phi}{\alpha}\left( \frac{
   M^2_{Z_2}}{m^2_{Z_1}} - 1 \right). 
\label{tzz}
\end{equation}
In our model the $S$ and $U$ parameters do not get contribution
from the $Z - Z'$ mixing~\cite{hold,rpd}. 

There are a few ways to get constraints on the 
mixing angle $\phi$ and the $Z^2$ mass. For example, 
a constraint on the $Z-Z'$ mixing can be obtained from the
$Z$-decay data. A bound for the mixing angle is~\cite{rhnm}
$ - 0.00018 \leq \phi\leq 0.00285$.

The total values of the  $S,\ T$ and $U$ parameters in this model 
are the sum of  the bilepton and the $Z-Z'$ contributions
\begin{eqnarray}
S_{\rm rhn}& = &   
 \frac{1}{4\pi} 
\Biggl\{
 5 \ln \frac{M_{+}^2}{M_0^2}
  + \frac{1}{3} 
  \left[
  13    \bar{F}_0(m_Z^2,M_+,M_+) -  \bar{F}_0(m_Z^2,M_0,M_0)
  \right]
\nonumber\\
&& \qquad
  + 4 \left[
      \bar{F}_3(m_Z^2,M_+,M_+) -   \bar{F}_3(m_Z^2,M_0,M_0)
  \right]
\Biggr\},\nonumber\\
 T_{\rm rhn}& = &   \frac{3\sqrt{2}G_F}{16\pi^2\alpha}
\left[ 
  M_+^2 + M_0^2 - \frac{2M_+^2M_0^2}{M_+^2-M_0^2}
  \ln \frac{M_+^2}{M_0^2}
\right] \nonumber \\
       &  & + \frac{1}{4\pi\ s^2_W}
\left[ 2    \bar{F}_0(0,M_+,M_0) +  \ t^2_W \ln\frac{M_+^2}{M_0^2}
\right]
+ \frac{\tan^2 \phi}{\alpha} \left( \frac{
M^2_{Z' }}{m^2_{Z  }} - 1 \right),
\label{ttot} \\
U_{\rm rhn} &=& - \frac{1}{\pi}
\Biggl\{ 2\left[  F_0(0,M_+,M_0) - \ s_W^2 
F_0(m^2_Z,M_+,M_+)\right]\nonumber\\
& & + \frac{2}{3} - \frac{(M^2_+ + M^2_0)}{2\ m_W^2}
  \left[  \bar{F}_0(m_W^2,M_+,M_0) - 
  \bar{F}_0(0,M_+,M_0)\right] \nonumber\\
& & - \frac{1}{4}\left[  \bar{F}_0(m_Z^2,M_+,M_+) +   
\bar{F}_0(m_Z^2,M_0,M_0) \right]
  + \frac{5}{2}   \bar{F}_0(m_W^2,M_+,M_0)\nonumber\\
& &   - 3 \left[  \bar{F}_3(m_Z^2,M_+,M_+) + 
  \bar{F}_3(m_Z^2,M_0,M_0) \right]\nonumber\\
& & + \frac{5}{m_W^2}  \left[  \bar{F}_4(m_W^2,M_+,M_0) - 
      \bar{F}_4(0,M_+,M_0)\right] - 6 \bar{F}_3(m_W^2,M_+,M_0)
\Biggr\}.\nonumber
\end{eqnarray}
In  Eq .~(\ref{ttot}) we have renamed the physical $Z_1$ and $Z_2$ 
to be usual $Z$ and $Z'$.

From~(\ref{Ffunction}),~(\ref{fd}) and~(\ref{f4})
it is easy to see that in the limit $M_+,\  M_0, \ M_{Z'} 
\rightarrow  \infty$   all  values $ \ S_{\rm rhn} ,\ T_{\rm rhn} ,\  
U_{\rm rhn} $ tend to zero in accord with 
the decoupling of heavy particles~\cite{ac}.  

With the help of~(\ref{masplit}) we can expand functions in $U_{\rm rhn}$
without any assumption in advance.
Most of the effects on precision measurements can be described by 
the three parameters  calculated above.

\section{ $\bf S,\ T,\ U$ parameters in the minimal 3 3 1 model}

\hspace*{0.5cm}Many useful details on the model are given 
in Ref.~\cite{dng}.
In~\cite{saken,frha} the parameters for the considered model 
are calculated without the $Z-Z'$ mixing contribution.
For our aim we note that Eqs.~(\ref{tphi}) and ~(\ref{masszw}) 
are  still correct. Therefore as in the above considered model the
$Z-Z'$ mixing gives contribution 
to the $T$ parameter  {\it only}, 
and the contribution is the same as in the model
with right-handed neutrinos
\begin{eqnarray}
 T_{\rm min}& = &   \frac{3\sqrt{2}G_F}{16\pi^2\alpha}
\left[ 
  M_{++}^2 + M_+^2 - \frac{2M_{++}^2M_+^2}{M_{++}^2-M_+^2}
  \ln \frac{M_{++}^2}{M_+^2}
\right] \nonumber \\
       &  & + \frac{1}{4\pi\ s^2_W}
\left[ 2   \bar{F}_0(0,M_{++},M_+) +  3 \ t^2_W \ln\frac{M_{++}^2}{M_+^2}
\right]
+ \frac{\tan^2 \phi}{\alpha} \left( \frac{
M^2_{Z'}}{m^2_{Z}} - 1 \right).
\label{ttotm} 
\end{eqnarray}

From Eqs.~(\ref{ttot}) and~(\ref{ttotm}) we see that the mixing 
contributions as expected~\cite{laluo} is positive, while the oblique
contributions can be negative in both versions of 331 models.
 
 The oblique contribution to the $S$ parameter is given in 
Ref.~\cite{frha}. However for the $U$ parameter one term was missed 
in the expression of $ U \vert_P $.
The correct expression for this part is
\begin{equation}
 U \vert_P = \frac{2}{\pi}
\Biggl[ 2 \ s_W^2   \bar{F}_0(m_Z^2,M_{++},M_{++}) - 
 \ s_W^2   \bar{F}_0(m_Z^2,M_{+},M_{+}) -
  \bar{F}_0(0,M_{++},M_+)\Biggr]\ .
\end{equation}

From Eq.~(\ref{ttotm}) we see that the $Z-Z'$ mixing contribution
increases by square of $Z'$ mass. Analysis in~\cite{dng} gives
$- 5\times 10^{-3} \leq \phi \leq 7 \times 10^{-4}$ from 
the low-energy experiment.
According to the recent analysis~\cite{pisano} the $Z'$ in this model
has very large lower limit  $M_{Z'}> 14$ TeV. With this mass,
the mixing contribution is valuable.
With $M_{Z'} = 1$ TeV the mixing contribution is about 4\%,
that's  why it was neglected in the previous analysis~\cite{frha}.

We note that results in this section  are correct for another 
3 3 1 version -- an 3 3 1 model with heavy charged lepton~\cite{plei}.
However, one point should be made here that the
condition~(\ref{masplit}) is correct for the mentioned 3 3 1 model with 
heavy charged lepton, but it is violated in the minimal version. 
\begin{figure} [htbp]
\begin{center}
\epsfile{file=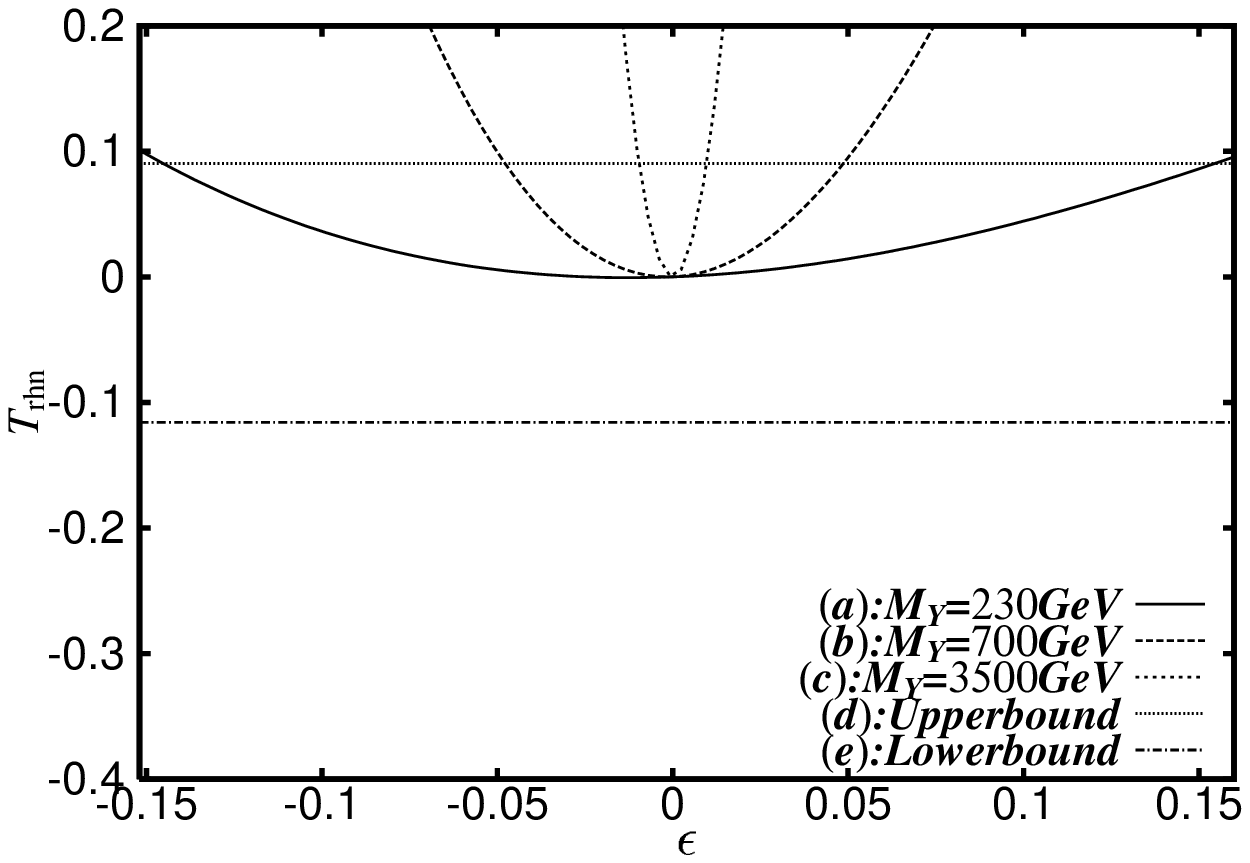,width=90mm,height=90mm}
\caption{$T_{\rm rhn}$ as functions of $\epsilon$ for three
values of $M_{Y^+}$:(a) $M_{Y^+}= 230$ GeV, (b) $M_{Y^+}= 700$ GeV,
and (c) $M_{Y^+}= 3500$ GeV. The horizontal lines (d) and (e) are
an upper and a lower limit on 
the experimental fit substracted the SM contribution $\Delta T_{\rm SM}$
for $m_H = 100$ GeV.}
\label{trhn}
\end{center}
\end{figure}

\section{Numerical evaluation}

\hspace*{0.5cm}For our initial purpose we consider the $\rho$ parameter 
 --  one of the most important quantities of the SM, having  a 
leading contribution in terms of the $T$ parameter
is very useful to get the new-physics effects (see,
for example~\cite{zprim,laluo,lis}). Defined at the zero point of momentum
$Q^2 = 0$ the $T$ parameter which is equivalent to $\Delta \rho$
has some advantage over the $U$ parameter  (to deal with 
F functions there, we have to suggest a  prior relationship between
bileptons masses and $m^2_Z,\ m^2_W$). 
Neglecting the $Z-Z'$ mixing
contribution which is  approximately  10 \% (for $\phi = 10^{-3}, 
M_{Z'}= 700$ GeV),  the $S,\ T$ parameters can be
rewriten in terms of two parameters  $\epsilon$ and $ \delta$  
as follows
\begin{eqnarray}
T_{\rm rhn} & = & \frac{1}{4\pi s^2_W c^2_W}
\left[\frac{3}{4}\frac{\epsilon^2}{\delta(M_+)} (2 - 3\epsilon + 2\epsilon^2) 
+ s^2_W  \epsilon  - \frac{\epsilon^2}{2} + O(\epsilon^4) \right],\nonumber\\
S_{\rm rhn} & = & \frac{1}{4\pi}
\left[ 5 \epsilon \left(1 - \frac{\epsilon}{2} + \frac{\epsilon^2}{3} 
\right) - \frac{2}{3} \delta(M_+) - \frac{\delta^2(M_+)}{15}
+ \frac{17}{90}\epsilon \delta(M_+)\right.\nonumber\\
& & \qquad \left. + \frac{5}{126}\epsilon \delta^2(M_+)
 +  O(\epsilon^4, \delta^3 (M_+))   \right],
\label{strhn}
\end{eqnarray}
where $\epsilon \equiv \varepsilon(M_+,M_0) = \frac{M_+^2 -
M_0^2}{M_0^2}$.

It is to be noted that, due to the mass splitting condition~(\ref{masplit}),
for given $M_+$, the parameter  $\epsilon$  is  bounded in the interval 
\begin{equation}
- \frac{m^2_W}{M^2_{+}} \leq  \epsilon  \leq \frac{m^2_W}{M^2_{+}},\
\epsilon  <  \delta( M_+) =  \frac{m^2_Z}{M^2_{+}}.   
\label{lim}
\end{equation}
For the heavier $M_+$, the interval of definition $\epsilon \in
\left[- \frac{m^2_W}{M^2_{+}},  \frac{m^2_W}{M^2_{+}} \right]$ becomes
shorter. With the interval of definition given by~(\ref{lim}), the $S$ 
and $T$ parameters are bounded too. In addition the
$T_{\rm rhn}$ is negative in the region $- \epsilon_C \le 
\epsilon \le 0$ where $\epsilon_C \simeq \frac{2 s_W^2}{3}\delta(M_+)$.

In  Fig.~\ref{trhn}   we plot the $T$   
parameter as function of the mass splitting parameter $\epsilon$
for the three choices $M_{Y^+} = 230,\  700$ and 3500 (GeV), respectively. 
The horizontal lines  are  experimental fit~\cite{rpd} after
substracting the SM contributions $\Delta T_{\rm SM}$~\cite{hag}
%for
%$m_H = 100$ GeV and 300 GeV, respectively,
\begin{eqnarray}
\Delta T_{\rm SM}& = &  + (0.130 - 0.003 x_H) x_t +0.003 x_t^2 
- 0.079 x_H -0.028 x^2_H \nonumber \\ 
& & \qquad  +0.0026 x^3_H, \nonumber\\
\end{eqnarray}
where $x_t$  and $x_H$ are  defined by
\begin{equation}
x_t \equiv \frac{m_t - 175~{\rm GeV}}{10~{\rm GeV}},\
x_H \equiv \log (m_H/100~{\rm GeV}). 
\label{eq:higgs_xh}
\end{equation}

 We choose the standard-model reference point at $m_t = 174$ GeV~\cite{rpd},
and  $m_H  = 100 $ GeV.
Fig.~\ref{trhn}  shows 
that $-0.0095  \stackrel{<}{\sim}  \epsilon  \stackrel{<}{\sim}  0.0096$ 
for $M_{Y^+} = 3500$ GeV,
 $-0.0475   \stackrel{<}{\sim} \epsilon  \stackrel{<}{\sim}  0.0483$  
for $M_{Y^+} = 700$ GeV, and
  $-0.144  \stackrel{<}{\sim} \epsilon  \stackrel{<}{\sim} 0.154$ 
for $M_{Y^+} = 230$ GeV.
This means that splitting in the bilepton masses is quite narrow
about 15 \% for the $M_+\sim 200 GeV$  , and decreases 
for the higher $M_{Y^+}$. 
This result is approximately consistent with the mass splitting given by 
the VEV structure~(\ref{first}).
\begin{figure} [htbp]
\begin{center}
\epsfile{file=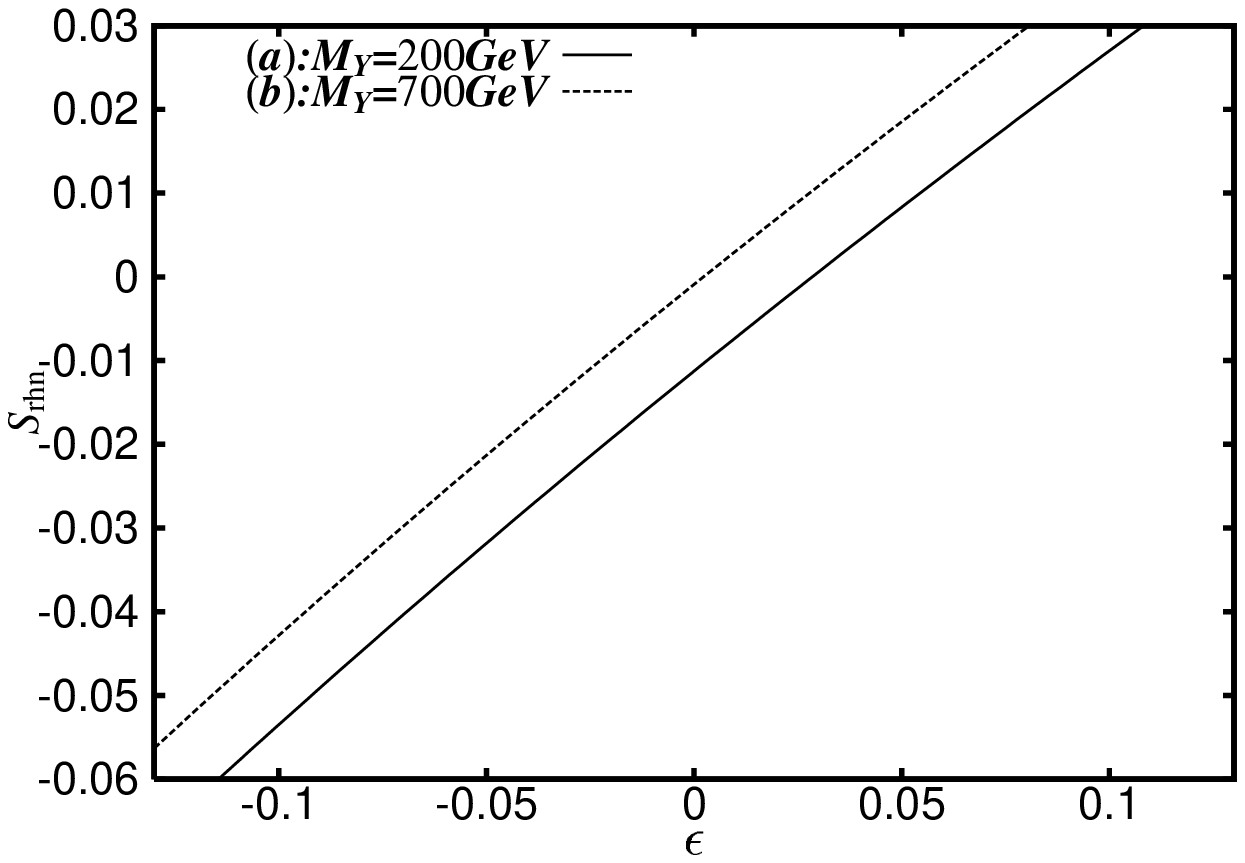,width=90mm,height=90mm}
\caption{$S_{\rm rhn}$ as functions of $\epsilon$ for two
values of $M_{Y^+}$:(a) $M_{Y^+}= 200$ GeV, (b) $M_{Y^+}= 700$ GeV.}
\label{srhn92}
\end{center}
\end{figure}

In  Fig.~\ref{srhn92}   we plot the $S$   
parameter as function of the mass splitting  parameter $\epsilon$
for the two choices $M_{Y^+} = 200$ GeV  and 700 GeV, respectively.
We see that the $S$ parameter  is  increasing function of 
the bilepton mass. However, due to decreasing of the definition
interval~(\ref{lim}), the running interval of the $S$ paremeter
becomes shorter too, eg: $- 0.018 \leq S \leq 0.05$
for $M_+ = 200$ GeV, while   $- 0.000082 \leq S \leq 0.0026$
for $M_+ = 1500$ GeV. This means that if experimental data
is closed to the SM zero point: $m_H  = 100 $
GeV, $m_t = 175$ GeV, the bilepton $X^0, Y^+$ will have large masses.
In this case the $Z-Z'$ mixing contribution has to be included. 
Thus we get a bound for the oblique  $S$ parameter:
$ - 0.06  \stackrel{<}{\sim} S  \stackrel{<}{\sim} 0.04$.
 
\begin{figure} [htbp]
\begin{center}
\epsfile{file=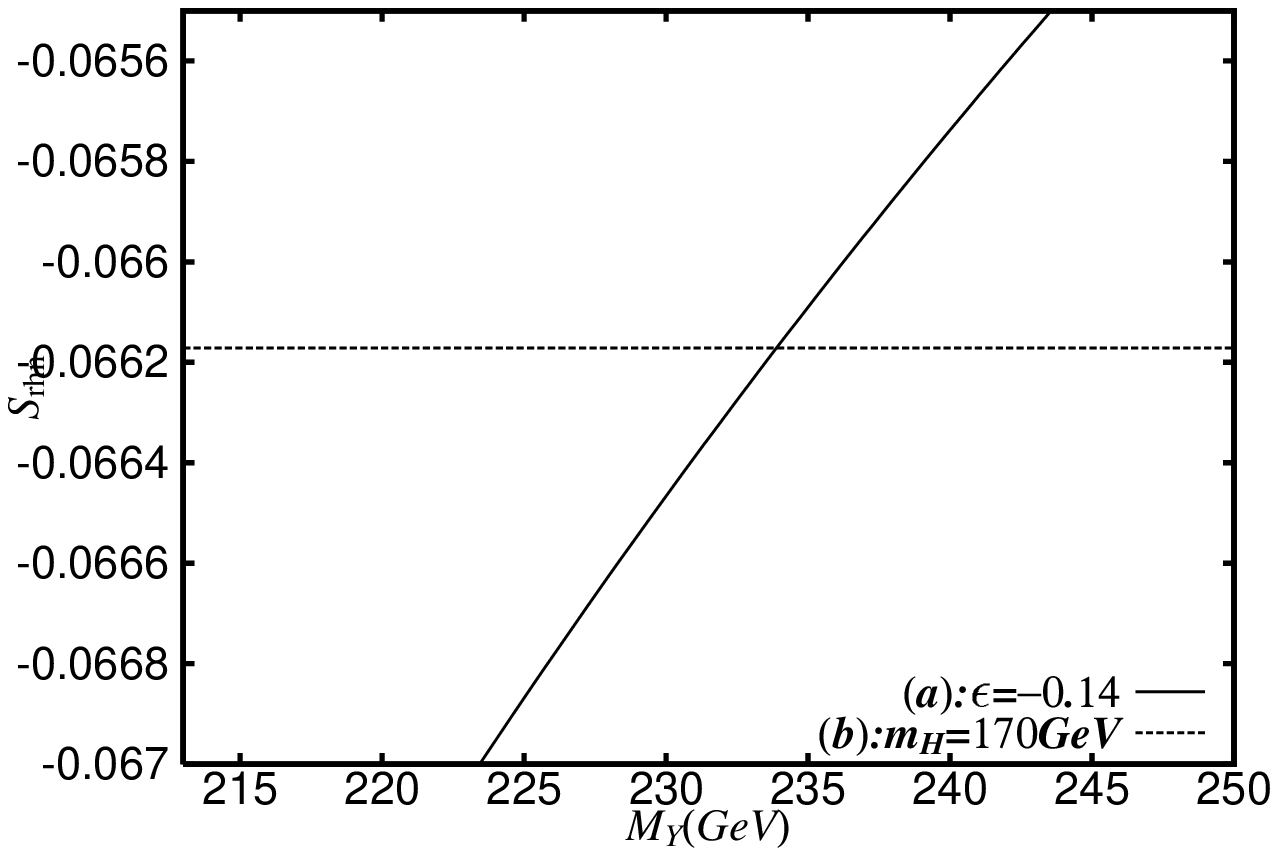,width=90mm,height=90mm}
\caption{$S_{\rm rhn}$ as functions of $M_{Y^+}$ for (a): 
$\epsilon = - 0.14$. The horizontal line (b) indicates an upper limit on
the experimental fit substracted the SM contribution $\Delta S_{\rm SM}$.}
\label{srhn}
\end{center}
\end{figure}

In  Fig.~\ref{srhn} we plot $S_{\rm rhn}$ as function of $M_+$ for
 (a): $\epsilon = - 0.14 $  as its maximum value for $M_0$ in the range
of  230 GeV. As before the horizontal line is a lower bound
on the experimental fit 
substracting the SM  contribution $\Delta S_{\rm SM}$~\cite{hag}
\begin{equation}
\Delta S_{\rm SM} =   - 0.007 x_t + 
 0.091 x_H - 0.010 x^2_H. \nonumber 
\end{equation}
This figure shows that  an allowed region for the  mass of the charged 
bilepton  $213 \leq M_{Y^+} \leq 234$ (GeV). It follows an allowed region 
for the mass of the neutral $X^0$: $230 \leq M_{X^0} \leq 251$ (GeV).

Note that the result is dependent at the top and Higgs masses.

\section{ Summary and conclusions}

\hspace*{0.5cm}In this paper we have calculated both the oblique 
and the mixing contributions to $S,\ T,\ U$ parameters.
The mixing contribution is negligible if mass of $Z'$
is less than  1 TeV, but it will be valuable for the
$Z'$ mass higher than 10 TeV.

We have shown that the oblique contributions to the $S$ and $T$
parameters are bounded, and can be negative. This result is 
interesting because
of the present experimental data seem to favor to negative
value for $T$. Since most of precision measurements can be 
described by the $S,\  T$ and  $U$ parameters
the obtained expresions  are very important
for the future data analysis.

We have mentioned that the bilepton mass splitting by the VEV structure
in the 3 3 1 model with r. h. neutrinos is smaller than those
in the minimal version. With this condition, we can get numerical
expression for the $U$ parameter without any assumption in advance. 

The oblique $S$ and $T$ parameters decrease with higher masses
of the bileptons. Thus in this case, the $Z-Z'$ mixing contribution
has to be considered.

As a consequence we  have found that the bilepton mass
splitting is quite narrow about 15 \% for
the singly-charged bilepton  mass around 200 GeV , 
and decreases for the higher
mass of the bileptons. Therefore in the future studies
it is acceptable  to put $M_{Y^+} \simeq M_{X^0}$.

From the Higgs structure and ``wrong '' muon decay we have got 
the first bound on the $M_{X^0}$: . 
The analysis based on~(\ref{strhn}) indicates that the neutral
bilepton is heavier, namely: for
  $213 \leq M_{Y^+} \leq 234$ (GeV), the allowed region
for  $M_{X^0}$: $230 \leq M_{X^0} \leq 251$ (GeV). 
% $M_{Y^+} \geq 260$ GeV and   $M_{X^0} \geq 280$ GeV.

As mentioned above, the constraints on bilepton masses are dependent
upon reference choices of the Higgs mass (even, on the top mass too).
Hence discovery of the Higgs particle will give a window to the new
particles in the SM extensions.
  
We hope to return to the data analysis in the future.

\section*{Acknowledgement}
The authors thank K. Sasaki for very helpful discussions.
The work of HNL is supported  by the JSPS grant No L98509.
He thanks Department of Physics, Chuo University 
for warm hospitality extended to him. 
This work is partially supported by Grant in Aid of
Ministry of Education, Science ans Culture, Priority area A, 
Priority area B (Supersymmetry and Unified Theory) and 
Basic research C.
\appendix
\renewcommand{\thesection}{Appendix {\Alph{section}}}
\section{}
\renewcommand{\thesection}{{\Alph{section}}}
\makeatletter
\setcounter{secnumdepth}{5}
\setcounter{tocdepth}{5}
\renewcommand{\theequation}{\thesection.\arabic{equation}}
\@addtoreset{equation}{section}
\makeatother

Functions  used in this paper are given in~\cite{frha}, however
we correct a misprint there
(in Eq. (A.1) below)
\begin{eqnarray}
\lefteqn{
  \bar{F}_0(s,M,m) = 
  \int^1_0 dx\, \ln \left( (1-x)M^2 + x m^2 - x(1-x)s \right)
  - \ln M m
}
\nonumber\\
   &=&
\left\{
\begin{array}{l}
  -\frac{2}{s} \sqrt{(M+m)^2-s} \sqrt{(M-m)^2-s}
  \ln \frac{ \sqrt{(M+m)^2-s} + \sqrt{(M-m)^2-s} }{ 2 \sqrt{Mm} }
\\
  \qquad
  + \frac{M^2-m^2}{s} \ln \frac{M}{m} - 2 \ ,
  \qquad \mbox{for} \quad s < (M-m)^2 \ ,
\\
  \frac{2}{s} \sqrt{(M+m)^2-s} \sqrt{s-(M-m)^2}
  \arctan \sqrt{\frac{s-(M-m)^2}{(M+m)^2-s}}
\\
  \qquad
  + \frac{M^2-m^2}{s} \ln \frac{M}{m} - 2 \ ,
  \qquad \mbox{for} \quad (M-m)^2 < s < (M+m)^2 \ ,
\\
  \frac{2}{s} \sqrt{s-(M+m)^2} \sqrt{s-(M-m)^2}
  \left[
    \ln \frac{ \sqrt{s-(M+m)^2} + \sqrt{s-(M-m)^2} }{ 2 \sqrt{Mm} }
    - i \pi
  \right]
\\
  \qquad
  + \frac{M^2-m^2}{s} \ln \frac{M}{m} - 2 \ ,
  \qquad \mbox{for} \quad (M+m)^2 < s \ .
\end{array}
\right.
\end{eqnarray}

\end{document}